\documentclass[pdflatex,sn-basic]{sn-jnl}
\usepackage{lmodern}
\usepackage{graphicx}%
\usepackage{epstopdf}
\usepackage{multirow}%
\usepackage{amsmath,amssymb,amsfonts}%
\usepackage{amsthm}%
\usepackage{mathrsfs}%
\usepackage[title]{appendix}%
\usepackage{xcolor}%
\usepackage{textcomp}%
\usepackage{manyfoot}%
\usepackage{booktabs}%
\usepackage{algorithm}%
\usepackage{algorithmicx}%
\usepackage{algpseudocode}%
\usepackage{listings}%
\usepackage{natbib}
\usepackage{subcaption}
\usepackage{fullpage}
\newcommand{\uinf}{U_{\infty}}
\newcommand{\meantau}{\overline{\tau_w}}

\newcommand{\ut}{u_{\tau}}

\begin{document}

\title[Article Title]{Signature of Pressure Gradient History on Wall Shear Stress in Turbulent Boundary Layers}

\author*[1]{\fnm{Marco} \sur{Mattei}}\email{mmattei2@illinois.edu}
\author[1]{\fnm{Theresa} \sur{Saxton-Fox}}\email{tsaxtonf@illinois.edu}

\affil*[1]{\orgdiv{Department of Aerospace Engineering}, \orgname{University of Illinois at Urbana Champaign}, \orgaddress{\street{104 S Wright St}, \city{Urbana}, \postcode{61801}, \state{Illinois}, \country{United States of America}}}

\abstract{
Our experiments evaluate the presence of history effects on wall shear stress in turbulent boundary layers (TBLs) subjected to five sequences of favorable and adverse pressure gradients (FAPGs) of increasing strength. Steady but spatially varying FAPGs are imposed on a flat plate using a deformable convex false ceiling. An array of three one-dimensional capacitive probes, positioned in the adverse pressure gradient (APG) region downstream of the favorable pressure gradient (FPG), enables direct and high-frequency measurements of wall shear stress. We first characterize the sensor performance under canonical zero pressure gradient (ZPG) conditions across a range of Reynolds numbers. With increasing FAPG strength, a non-monotonic increase in skin friction and a reduction in normalized wall shear stress fluctuations are  observed, indicating long-lasting upstream FPG influence within the downstream APG region. These effects are more pronounced at lower Reynolds numbers. Additionally, the influence of large-scale structures in FAPG on wall shear stress is studied through coherence spectrum, two points cross-correlation, and convection velocity. 
}


\maketitle

\section{Introduction}\label{sec_intro}
Despite the extensive focus on canonical turbulent boundary layers (TBLs) under zero pressure gradient (ZPG) conditions, most engineering flows are impacted by streamwise pressure gradients (PGs), which arise from surface curvature, geometric effects, or external flow interactions \citep{vishwanathan2023history, preskett2025effects}. TBLs subjected to favorable and adverse pressure gradients (FPG and APG) deviate from canonical ZPG flows, posing significant challenges in both computational and experimental contexts. Accurate prediction of flow development and surface forces for non-canonical boundary layers remains a critical challenge. Pressure gradient alters the balance among turbulence production, transport, and dissipation, the structure of coherent motions, the skin friction drag, and the tendency of the flow to separate.  The boundary layer does not respond uniformly across all scales but different structures exhibit varying behavior depending on their wall-normal location. The effects of single-sign PGs, either favorable or adverse, immediately downstream of a ZPG have been extensively studied \citep{clauser1954turbulent, perry1966velocity, kline1967structure, schloemer1967effects, smits1983low, nagano1993effects, maciel2006self, lee2009structures, monty2011parametric, joshi2014effect, yuan2014estimation, kitsios2016direct, wang2020effects, pargal2022adverse, pozuelo2022adverse, romero2022properties, deshpande2023reynolds, volino2025outer}. A FPG accelerates near-wall structures, leading to increased wall shear stress and a reduced frequency of bursting events and Reynolds stresses, which in turn results in lower turbulence levels. The mean velocity profile thickens in the viscous sublayer and buffer region, while the wake region becomes thinner. Depending on the pressure gradient strength and the freestream conditions, the TBL may relaminarize.  Conversely, an APG causes boundary layer growth and deceleration, leading to reduced wall shear stress, intensified bursting accompanied by increased turbulence production, and a secondary peak in Reynolds shear stresses in the outer region. Under strong APG conditions, backflow may occur near the wall, potentially leading to flow separation.  

When spatially varying pressure gradients are introduced, flow development becomes more complex. These configurations exhibit a spatially dependent Clauser pressure gradient parameter ($\beta = \frac{\delta^*}{\tau_w}\frac{dP}{dx}$), are not well understood, are prone to flow history effects \citep{bobke2017history, fritsch2022turbulence, fritsch2022fluctuating}, and the boundary layer exhibits a non-uniform response to changes in the force balance across scales. While the near wall region ($y^+<10$) responds almost instantaneously, the mean flow in the outer region adjusts with a certain delay \citep{gungor2024turbulent}. The local state of a boundary layer depends not only on the local pressure force, but also on its streamwise evolution, particularly its rate of change, $d\beta/dx$, often referred to as a local disequilibrating effect \citep{gungor2024effect}, as well as on the cumulative upstream flow history \citep{spalart1993experimental, vinuesa2017revisiting, devenport2022equilibrium}. Even when local flow quantities such as friction Reynolds number ($Re_{\tau}$) and $\beta$ are matched, different upstream evolutions of $\beta$ can lead to distinct velocity profiles and turbulence statistics \citep{sanmiguel2017adverse, preskett2025effects}. 

In FAPG sequences, the flow behavior in the downstream APG region differs significantly from that of a TBL initially developed under a ZPG \citep{tsuji1976turbulent, volino2020non, williams2020experimental, parthasarathy2023family}. Following a FPG, as part of an FAPG sequence, the outer region of the TBL remains frozen \citep{baskaran1987turbulent} and does not exhibit the outer region Reynolds stress peak typically observed in APG flows \citep{lee2008effects}. Additionally, relaminarized TBLs induced by strong FPGs are less prone to separation under subsequent APGs \citep{Balin2021DirectGradients}. \cite{parthasarathy2023family} recently analyzed several TBLs subjected to FAPG sequences of varying magnitude and showed that the upstream FPG exerts a pronounced and lasting downstream influence, even after the onset of the APG. Their results revealed that the response of the TBL in the APG region was driven by an internal layer that grew to cover about 20\% of the TBL thickness by the end of their field of view, while the outer layer developed in the APG region with minimal changes to the state established by the preceding FPG. 

Wall shear stress represents a physically meaningful quantity for assessing the response of a TBL to upstream PGs and evaluating history effects. In statistically stationary flows, it is usually decomposed as $\tau_w(\mathbf{x}, t) =\meantau(\mathbf{x})+\tau_w'(\mathbf{x},t)$, where the time-averaged term reflects the overall flow conditions over a surface, whereas the fluctuating part captures unsteady momentum transfer to the wall linked to structural fatigue and fluid-induced noise \citep{Haritonidis1989WSS, Colella2003}. Wall shear stress is used in the definition of the friction velocity ($\ut=\sqrt{\meantau/\rho}$) and viscous length scales ($l^{+} = \nu / \ut$), and also serves as an indicator of important flow physics, including laminar to turbulent transition, flow separation, and reattachment. 
Despite its importance, wall shear stress under streamwise PG sequences remains relatively underexplored primarily due to limitations in measurement techniques. The majority of existing measurement methods rely on strong assumptions about the boundary layer structure or suffer from limited spatial resolution, intrusiveness, and lack of high-frequency capability. Recent work demonstrated the influence of PG history on skin friction using oil-film interferometry \citep{virgilio2025pressure}. In parallel, advances in MEMS-based capacitive shear stress sensors \citep{Pabon2018CharacteristicsMeasurements, Mills2021TemperatureTunnels} offer direct, high-frequency, and non-intrusive measurements with directional sensitivity, making them well suited for capturing both the mean and unsteady wall response in TBLs subjected to FAPG sequences.

Our study aims to provide new insights into the physics of history effects induced by spatially varying PGs in turbulent wall-bounded flows, with the goal of advancing the understanding of the mechanisms governing wall turbulence and supporting the development of improved computational models. We employ three capacitive MEMS-based sensors (DirectShear\textsuperscript{TM} Model CS-0210, IC2) mounted on a flat plate to obtain direct and high-frequency measurements of wall shear stress. Sensor performance is first quantified under ZPG conditions by comparing mean and fluctuating terms against reference data from canonical TBLs. Although these sensors have been thoroughly described and characterized in prior work \citep{PabonThesis, Mills2019CharacterizationTunnels, Freidkes2020ASensor}, a systematic evaluation of their response across a range of Reynolds numbers has not, to our knowledge, been reported. We then investigate TBLs subjected to steady but spatially varying sequences of FAPG of different strength, induced on the flat plate through controlled deformation of a false ceiling of the test section. Direct measurements are conducted within the APG region downstream of the FPG. By analyzing both mean and fluctuating components, we evaluate whether and how the PG evolution influences the wall shear stress response. In analogy to the delayed response of the boundary layer structure to pressure variations \citep{vishwanathan2023history, gungor2024turbulent}, we expect similar delays in both mean and fluctuating components of wall shear stress. By directly measuring $\tau_w(\mathbf{x}, t)$, we capture the integrated response of the TBL, while also resolving the signature of larger scale motions. These structures are known to significantly contribute to the skin friction \citep{deck2014large}, influence both mean flow and near-wall turbulence through amplitude modulation and superposition effects \citep{na1998structure, abe2004very, hutchins2007evidence, hutchins2007large, harun2013pressure, preskett2025effects}, and induce distinct wall signatures in shear stress, pressure, and velocity-shear correlations \citep{thomas1983role,  marusic2001role, o2009chasing, luhar2013wall, fritsch2022fluctuating}. The influence of FAPG sequences on their dynamics can be inferred from coherence spectrum and two-point correlations of wall shear stress signals. Based on coherence results, we introduce a two-steps low-pass filter to isolate lower-frequency motions under different FAPG conditions without any flow assumptions. 

The structure of the manuscript is as follows. Section \ref{sec_exps} describes the experimental framework, the wall shear stress measurement array, and associated limitations and uncertainties. Section \ref{sec_res} presents Reynolds number effects on wall shear stress under ZPG conditions, followed by an analysis of flow history effects on mean and fluctuating skin friction and their influence on large-scale dynamics. Finally, section \ref{sec_conc} provides a summary and concluding remarks. 

\section{Experiments}\label{sec_exps}
\subsection{Boundary Layer Wind Tunnel}\label{BL_facility}
Experiments are conducted within the open circuit Boundary Layer Wind Tunnel (BLWT) at the Aerodynamics Research Laboratory of the University of Illinois at Urbana-Champaign. The flow undergoes conditioning in a settling chamber equipped with a 10.19 cm thick expanded aluminum honeycomb straightener and four 24-mesh stainless steel turbulence-reducing screens. It is then accelerated through an area contraction ratio of 27:1 before reaching the test section. The range of speeds that can be achieved goes from 1 to 40 m/s with maximum freestream turbulence intensity of 0.8\%. The dimensions of the test section are 0.381 m  $\times$ 0.381 m $\times$ 3.657 m. The TBL is generated across a 2.54 cm thick flat plate that runs through the entire test section and is installed at mid-height. The leading edge of the flat plate features a modified super-ellipse shape, and a 5 cm wide sandpaper strip with a grit size of 40 is installed on its top surface to trip the boundary layer to turbulent. The trailing edge of the flat plate is completed by a 0.61 m long angle-adjustable flap, used to fix the stagnation point of the flat plate flow at the leading edge. Further details on the wind tunnel can be found in \citep{Rodriguez2020}. 

\subsection{Flow conditions} 
\begin{table}[htbp]
\centering
\caption{Experimental flow conditions.}
\label{tab_flow_conds}
{\small
\begin{tabular}{c c c c c c c}
\toprule
Case & $\uinf$ (m/s) & $\delta$ (mm) & $T_{\mathrm{FAPG}}$ (ms) & $Re_x$ & $Re_{\tau}$ & $Re_{\theta}$ \\
\midrule
1 & 7.6  & 53 & 60 & $1.2\times 10^6$ & 1090 & 3386 \\
2 & 12.0 & 48 & 38 & $1.8\times 10^6$ & 1500 & 4804 \\
3 & 16.8 & 45 & 27 & $2.5\times 10^6$ & 1899 & 6222 \\
4 & 20.2 & 44 & 23 & $3.0\times 10^6$ & 2160 & 7166 \\
5 & 26.0 & 42 & 18 & $3.9\times 10^6$ & 2578 & 8699 \\
\bottomrule
\end{tabular}
}
\end{table}

Table \ref{tab_flow_conds} summarizes the experimental flow conditions, including the freestream velocity ($\uinf$), boundary layer thickness ($\delta$) at the start of the sensor array under incoming ZPG conditions, and the characteristic advection time across the FAPG region ($T_{\mathrm{FAPG}} = L / U_{\infty}$), where $L=0.457$ m is the length of the deformable ceiling. Also reported are the Reynolds number based on the distance from the leading edge of the flat plate ($Re_{x}$), friction Reynolds number ($Re_{\tau})$, and momentum Reynolds number ($Re_{\theta}$). For each freestream condition, a baseline ZPG case ($\beta=0$) is tested, followed by five non-equilibrium FAPG sequences of increasing strength, each characterized by ($d\beta/dx\neq0$). 

\begin{figure}[htbp]
  \centering
  \includegraphics[width=0.3\textwidth]{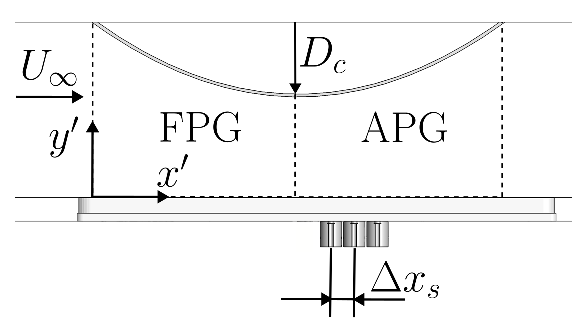}
  \caption{Schematic of the test region with deflected ceiling (not to scale) and sensors installed in the APG.}
  \label{Fig_schem_WT}
\end{figure}

\begin{figure}[htbp]
    \centering
    \begin{subfigure}[t]{0.35\textwidth}
        \centering
        \includegraphics[width=\textwidth]{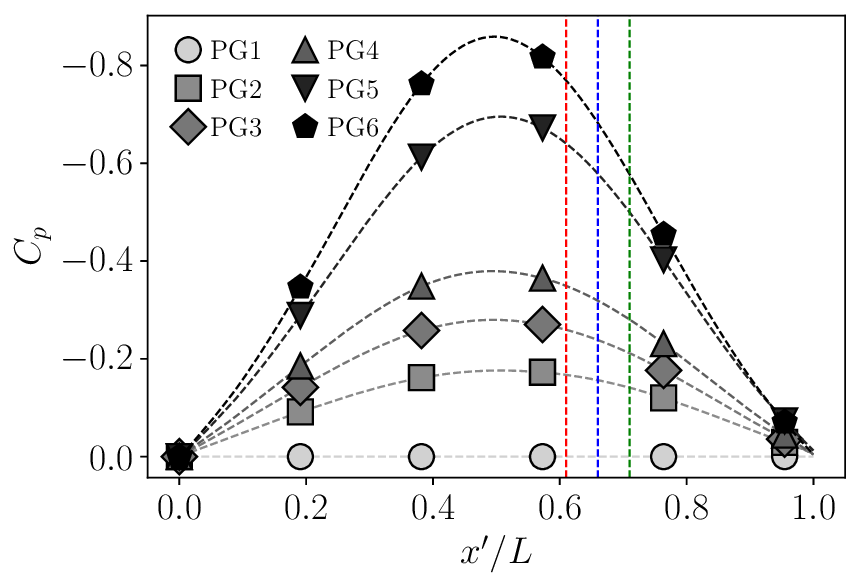}
        \caption{}
        \label{Fig_2}
    \end{subfigure}
    \hspace{0.04\textwidth}
    \begin{subfigure}[t]{0.35\textwidth}
        \centering
        \includegraphics[width=\textwidth]{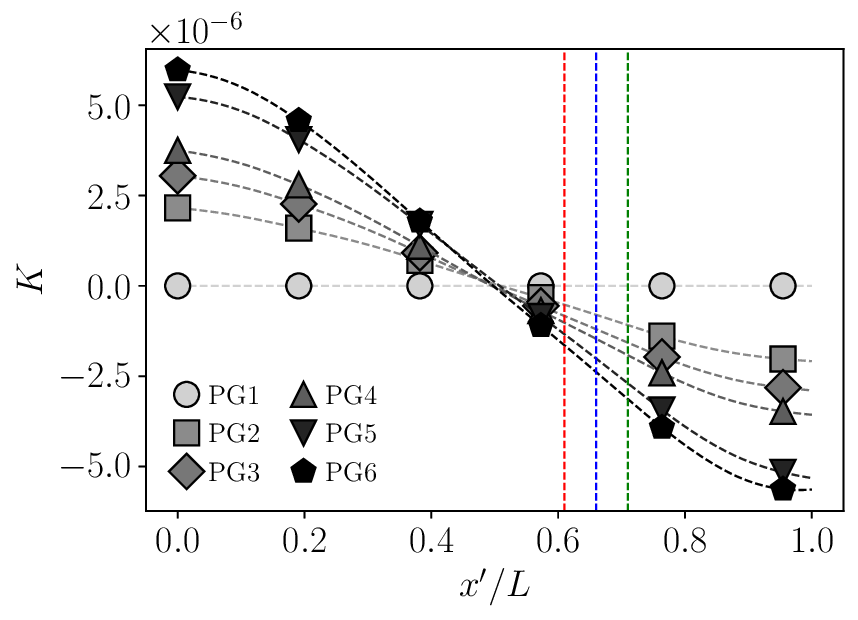}
        \caption{}
        \label{Fig_3}
    \end{subfigure}
    \caption{Spatial evolution of (a) pressure coefficient ($C_p$) and (b) acceleration parameter ($K$) for different pressure gradient sequences \citep{parthasarathy2022novel}.}
    \label{Fig_Cp_K}
\end{figure}

Figure \ref{Fig_schem_WT} shows a schematic of the 0.61 m long test region, which the TBL reaches after developing over the flat plate for 2.35 m. Here, FAPG sequences are imposed via a deformable convex ceiling \citep{parthasarathy2022novel}. Increasing the ceiling deflection ($D_c$) progressively strengthens both the FPG and APG regions beneath. The setup geometry ensures a constant pressure distribution across Reynolds numbers and decouples PG effects from surface curvature, enabling the study of identical pressure distribution under different flow conditions. This is critical for isolating flow history effects from Reynolds number effects \citep{gungor2024turbulent}. The sequential PGs are characterized by the pressure coefficient ($C_p$) and the acceleration parameter ($K= \frac{\nu}{U^2}\frac{dU}{dx}$) shown in figure \ref{Fig_Cp_K} as functions of the normalized streamwise coordinate $(x'/L)$, where $x'$ is measured from the start of the ceiling. Although wall shear stress is measured in the APG region, the sensors are located just downstream of the acceleration zone, and $K$ properly describes the local flow conditions. To account for the upstream development of the flow, which significantly influences the TBL response \citep{vinuesa2014experiments, parthasarathy2023family}, we report the spatially averaged acceleration parameter ($\overline{K}$), computed from $x'/L=0$ to each sensor location. These sensor locations are shown in figure \ref{Fig_Cp_K} with the dashed lines: red at $x'/L = 0.61$, blue at $x'/L = 0.66$, green at $x'/L = 0.72$, and the corresponding $\overline{K}$ values are reported in table \ref{tab:pg_conds}.

\begin{table}[htbp]
\centering
\caption{FAPG sequence parameters.}
\label{tab:pg_conds}
\begin{tabular}{c c c c c c}
\toprule
Test ID & $C_{p,\min}$ & $K_{\max} \times 10^{6}$ & $\overline{K}(x'/L = 0.61) \times 10^{6}$ & $\overline{K}(x'/L = 0.66) \times 10^{6}$ & $\overline{K}(x'/L = 0.72) \times 10^{6}$ \\
\midrule
PG1 & 0.00  & 0.00 & 0.00 & 0.00 & 0.00 \\
PG2 & -0.18 & 2.16 & 1.00 & 0.88 & 0.72 \\
PG3 & -0.28 & 3.04 & 1.37 & 1.20 & 0.98 \\
PG4 & -0.38 & 3.75 & 1.68 & 1.46 & 1.19 \\
PG5 & -0.70 & 5.23 & 2.49 & 2.17 & 1.79 \\
PG6 & -0.86 & 5.97 & 2.74 & 2.38 & 1.95 \\
\bottomrule
\end{tabular}
\end{table}

\subsection{Wall shear stress measurements and their limitations} 
\begin{table}[htbp]
\centering
\caption{Measurement parameters for different freestream conditions under ZPG flow. \textsuperscript{\#}Values estimated from linear extrapolation of friction velocity.} 
\label{tab:combined_params}
\begin{tabular}{c c c c c c c c c c c}
\toprule
Case & $d^*$ & $2d^*$ & $T_s$ (s) & $f_o^*$ & $f^+$ & $f_i^*$ & $l_1^+$ & $l_2^+$ & $l_d^+$ & $f_t/f_s$ \\
\midrule
1 & 0.475 & 0.950 & 128 & 89 & 2.64 & 1960 & 22.3 & 4.5  & 11.2 & 0.208 \\
2 & 0.521 & 1.042 & 64  & 52 & 0.73 & 1190 & 33.6 & 6.7  & 16.8 & 0.468 \\
3 & 0.557 & 1.114 & 46  & 35 & 0.40 & 822  & 45.6 & 9.1  & 22.8 & 0.859 \\
4 & 0.578 & 1.156 & 38  & 28 & 0.25 & 626  & 53.8 & 10.8 & 26.8 & 1.250 \\
5 & 0.608 & 1.215 & 26  & 20 & 0.15\textsuperscript{\#} & 501\textsuperscript{\#} & 67.5 & 13.5 & 33.7 & 1.789 \\
\bottomrule
\end{tabular}
\end{table}

\begin{figure}[htbp]
  \centering
  \includegraphics[width=0.16\textwidth]{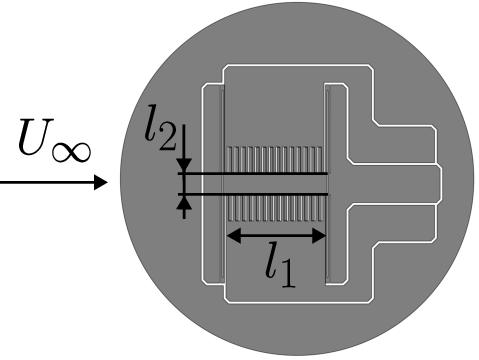}
  \caption{Schematic of sensing element (not to scale) \citep{ic2manual}.}
  \label{Fig_scheme_IC2}
\end{figure}

In this study, we present the results from three flush-mounted capacitive shear stress sensors (DirectShear\textsuperscript{TM} Model CS-0210) installed in the APG region, as visible in figure \ref{Fig_schem_WT}. The sensors are uniformly spaced at $\Delta x_s = 25$ mm,  aligned with the streamwise direction at the centerline of the test section and are installed at $x'/L=(0.61, 0.66, 0.72)$. These positions are selected by accounting for the physical dimensions of the sensors, the need to position them upstream of the separated flow region, which occurs on the ceiling at $x'/L\approx80\%$ \citep{parthasarathy2022novel}, and the requirement of minimizing their spacing for two-point correlation analysis. To ensure accurate flush-mounting, the sensor heads were installed with stringent tolerances. The maximum step between the sensor head and the flat plate has been measured at 0.026 mm, using a wall-normal dial gauge with a resolution of 0.013 mm, corresponding to 1.8 viscous units at the highest freestream speed. This value is below the hydrodynamic smoothness threshold of $\sim3.5-5$ wall units, where roughness effects are negligible \citep{shockling2006, HutchinsHW2009}.

Important experimental parameters evaluated at the start of the sensor array are summarized in Table \ref{tab:combined_params}. The normalized sensor spacing ($d^* = \kappa \Delta  x_s / \delta$, where $\kappa=1,2$ is an integer indicating the sensor pair separation) quantifies the streamwise distance covered by the sensor array, which ranges from $0.475\delta$ to $1.215\delta$ for friction Reynolds numbers from 1090 to 2578. The sampling time ($T_s$) varies with the freestream velocity and is chosen to ensure at least 15,000 boundary-layer turnover times  ($T_s U_{\infty} / \delta$) \citep{Mathis2009, Scarano2022}, in order to guarantee convergence of the flow statistics associated with the large-scale structures that modulate the near-wall turbulence \citep{OrluSchlatter2011}.  Table \ref{tab:combined_params} also reports the normalized frequencies in outer units ($f^*_o = f_s \delta / U_{\infty}$), inner units ($f^*_i = f_s \delta / u_{\tau}$), and the non-dimensional sampling frequency ($f^+ = 1 / \Delta t^+$, where $\Delta t^+ = \Delta t u_{\tau}^2 / \nu$)  with $f_s = 12.8$ kHz. These quantities are normalized using $\delta$ and $u_{\tau}$ values obtained from PIV measurements within the same experimental setup at freestream velocities $U_{\infty}=(7.6, 12.0, 16.8, 20.2)$ m/s \citep{DatabasePIV2023}.  For $U_{\infty}=26.0$  m/s, values marked with $^{\#}$ in table \ref{tab:combined_params} are calculated using a linearly estimated friction velocity based on experimental data for lower speeds. 

Figure \ref{Fig_scheme_IC2} displays a schematic of the sensing element, which has dimensions $l_1=1$ mm $\times$ $l_2=0.2$ mm, enabling one-directional measurements up to 300 Pa in the freestream direction. Little to no literature addresses the challenges of capacitive shear stress sensors in TBLs, particularly regarding Reynolds number effects on their performance. Some of these challenges can be addressed by an analogy to well-established techniques commonly used in wall-bounded flows. Extensive work has explored the spatial and temporal resolutions of thermal anemometry \citep{JohanssonAlfredsson1983, sharma1984, Ligrani1987, HutchinsHW2009, Hultmark2010, Smits2011}. For hot wire and hot film probes, attenuation is minimized when the characteristic dimension of the probe is sufficiently small to resolve the energetic scales and when the system is sufficiently fast to capture flow structures as they move across the sensor at a specific convection velocity \citep{HutchinsFreq2015}. Adequate spatial resolution is typically achieved when the viscous-scaled length of the wire ($l^+$) is below 10-20 and the length-to-diameter ratio exceeds 200 \citep{Ligrani1987, Alfredsson1988}. Additionally, the sampling frequency must be sufficiently high to prevent aliasing. Following this analogy, critical parameters defining the performance of the capacitive shear sensors are the viscous-scaled sensor lengths ($l_1^+,l_2^+, l_d^+=\sqrt{4l_1l_2/\pi}$) and the ratio of the highest frequency associated with the smallest scales within the turbulent boundary layer ($f_t$) \citep{HutchinsHW2009} to the sampling frequency ($f_s$). These quantities are listed in table \ref{tab:combined_params}. Based on the sensing element dimensions, the 1 mm size corresponds to $l_1^+$ values from 12.1 to 59.1, the 0.2 mm results in $l_2^+$ from 2.4 to 11.8. Assuming a characteristic diameter of 0.5 mm, $l_d^+$ yields values between 6.1 and 29.6.  
As the Reynolds number increases, spatial and temporal attenuation are expected to become more significant, similar to thermal anemometry techniques. Low levels of attenuation are anticipated for $l_1^+\lessapprox20-25$ and for sampling frequency at least twice the frequency associated with the smallest scales at $Re_{\tau}\leq1500$. However, we expect the measured magnitude of the skin-friction fluctuations to decrease for increasing sensor scaled dimension, in analogy to the observation for hot-film sensors \citep{DelAlamo2009EstimationApproximation, Hutchins2011Three-dimensionalLayer}.

\subsection{Measurement Uncertainty}\label{sec_unc}
\begin{figure}[htbp]
    \centering
    \includegraphics[width=0.32\textwidth]{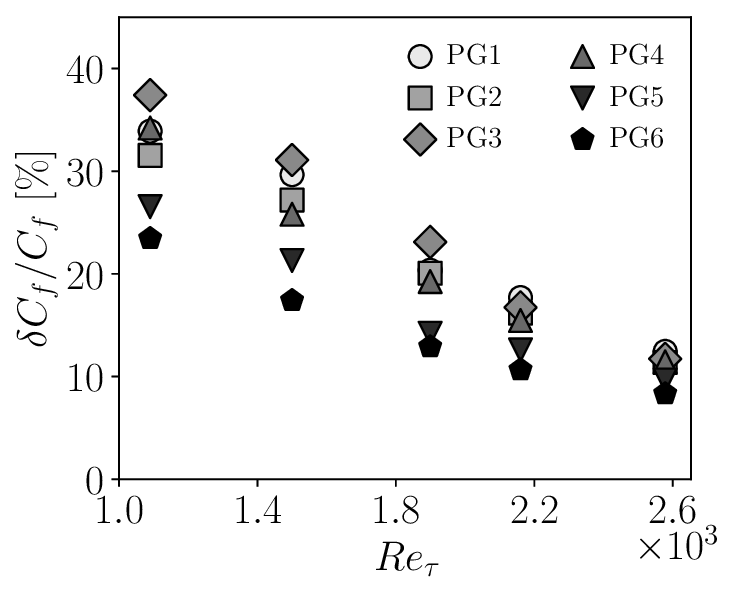}
    \caption{Relative uncertainty of the skin friction coefficient ($\delta C_f / C_f$) as function of $Re_\tau$ and different FAPG strengths for the first sensor.}
    \label{Fig_delta_Cf_ratio}
\end{figure}

The measurement uncertainty is quantified explicitly for the first sensor in the probe array as a representative case. The total uncertainty in skin friction coefficient ($C_f=\tau_w / (0.5 \rho U_\infty^2)$) is expressed as the root-sum-square of four primary contributions:
\begin{equation}
    \delta C_f = \dfrac{2\delta\tau_w}{\rho U_{\infty}^2} = \dfrac{2}{\rho U_{\infty}^2} \sqrt{\mathrm{(Res)}^2+ \mathrm{(Noise)}^2 +(DC_{\mathrm{acc}})^2 + \mathrm{(Temp)}^2}  \label{eq_unc_tau}
\end{equation}
Here, $\mathrm{Res}$ is the sensor resolution (1 to 1.2 mPa), the term $\mathrm{Noise}$ accounts for the electronic noise floor and is estimated as $n_f \times \sqrt{f_b} \times \frac{1}{S}$, where $n_f=2.4 \mu \mathrm{V}/\sqrt{\mathrm{Hz}}$, $f_b=5\ \mathrm{kHz}$ is the bandwidth frequency, and $S=1.99\ \mathrm{mV/Pa}$ is the dynamic sensitivity of the sensor. The DC measurement accuracy ($DC_{\mathrm{acc}}$) is evaluated individually for each acquisition. While the sensor datasheet specifies $DC_{\mathrm{acc}}$ as $\pm0.02\%\mathrm{FS}= \pm(6.0\times10^{-2})$ Pa over a 10-minute interval, we measured a $DC_{\mathrm{acc}}=2.2\times10^{-4}$Pa across the longer acquisition time. The last term ($\mathrm{Temp}$) accounts for offset drift due to temperature variation and is defined as 0.09\% FS per degree Celsius. Based on these considerations, the total measurement uncertainty in the wall shear stress is $\delta\tau_w = 0.073$ Pa at the lowest freestream condition. Similarly, the uncertainty in the normalized fluctuation magnitude of wall shear stress ($\tau_w'^+ =\tau'_{w,\mathrm{rms}}/\overline{\tau_w}$) is evaluated through:
\begin{equation}
   \dfrac{\delta \tau'^+_w}{\tau'^+_w} = \sqrt{\bigg(\dfrac{\delta \tau'^+_{w,\mathrm{rms}}}{\tau'^+_{w,\mathrm{rms}}} \bigg)^2 + \bigg(\dfrac{\delta {\tau_w}}{\tau_w} \bigg)^2} \approx \dfrac{\delta\tau_w}{\tau_w}
   \label{eq_unc_tau_plus}
\end{equation}
because $\delta(\mathrm{rms}) = \sigma/\sqrt{2N}$ \citep{bendat2011random}, and $N \gg 1$, the first term becomes negligible. Table \ref{tab:cf_uncertainty} reports the relative measurement uncertainty across different freestream conditions and sensor positions. Figure \ref{Fig_delta_Cf_ratio} shows the relative uncertainty of the skin friction coefficient ($\delta C_f/C_f$) as a function of $Re_\tau$ for different FAPG sequences and evaluated for the sensor positioned at $x'/L=0.61$. At the lowest freestream condition ($Re_\tau = 1090$), a strong dependence on the FAPG strength is observed. The relative uncertainty exceeds 35\% for PG2 and decreases to approximately 20\% for PG4--PG6. As $Re_\tau$ increases, $\delta C_f/C_f$ decreases and  differences across FAPG sequences diminish. At the highest Reynolds number analyzed ($Re_\tau = 2578$) , uncertainties converge to approximately 10\% for all cases. To evaluate repeatability of the measurements,  three independent acquisitions are performed at each flow condition. The normalized range across different individual measurements, defined as $(\max(C_f) - \min(C_f)) / \overline{C_f}$, decreases from approximately 15\% at the lowest freestream condition to 2.4\% at the highest, remaining below the measurement uncertainty range. 

\begin{table}[htbp]
\centering
\caption{Relative uncertainty for $\delta C_f / C_f\ [\%]$ and  $\delta\tau'^+_w / \tau'^+_w\ [\%]$ across different FAPG sequences for various $Re_{\tau}$ and sensor positions $x'/L$.}
\label{tab:cf_uncertainty}
\begin{tabular}{llcccccc}
\toprule
$Re_\tau$ & $x'/L$ & $\mathrm{PG1}$ & $\mathrm{PG2}$ & $\mathrm{PG3}$ & $\mathrm{PG4}$ & $\mathrm{PG5}$ & $\mathrm{PG6}$ \\
\midrule
1090 & 0.61 & 33.9 & 31.5 & 37.4 & 34.2 & 26.6 & 23.5 \\
1090 & 0.66 & 28.2 & 26.4 & 32.2 & 27.4 & 28.1 & 27.4 \\
1090 & 0.72 & 27.9 & 34.7 & 31.0 & 21.0 & 20.5 & 19.8 \\
1500 & 0.61 & 29.7 & 27.2 & 31.1 & 25.8 & 21.3 & 17.4 \\
1500& 0.66 & 21.9 & 21.2 & 23.2 & 21.7 & 20.9 & 19.5 \\
1500 & 0.72 & 20.8 & 26.6 & 23.0 & 18.6 & 16.1 & 16.4 \\
1899 & 0.61 & 20.4 & 20.1 & 23.1 & 19.3 & 14.2 & 12.9 \\
1899 & 0.66 & 17.3 & 16.3 & 16.6 & 16.2 & 14.0 & 13.8 \\
1899 & 0.72 & 17.4 & 19.2 & 16.1 & 13.8 & 13.2 & 12.9 \\
2160 & 0.61 & 17.7 & 16.2 & 16.7 & 15.5 & 12.6 & 10.7 \\
2160 & 0.66 & 13.4 & 14.0 & 14.2 & 13.1 & 11.6 & 11.8 \\
2160 & 0.72 & 14.0 & 14.0 & 13.8 & 11.9 & 10.8 & 11.4 \\
2578 & 0.61 & 12.5 & 11.4 & 11.7 & 11.4 & 9.7 & 8.3 \\
2578 & 0.66 & 10.6 & 10.0 & 10.6 & 9.8 & 8.8 & 8.7 \\
2578 & 0.72 & 10.5 & 10.3 & 9.9 & 9.4 & 8.7 & 8.6 \\
\bottomrule
\end{tabular}
\end{table}

\section{Results}\label{sec_res}

\subsection{Reynolds number effects under zero pressure gradient conditions}
\begin{figure}[htbp]
    \centering
    \begin{subfigure}[t]{0.32\textwidth}
        \centering
        \includegraphics[width=\textwidth]{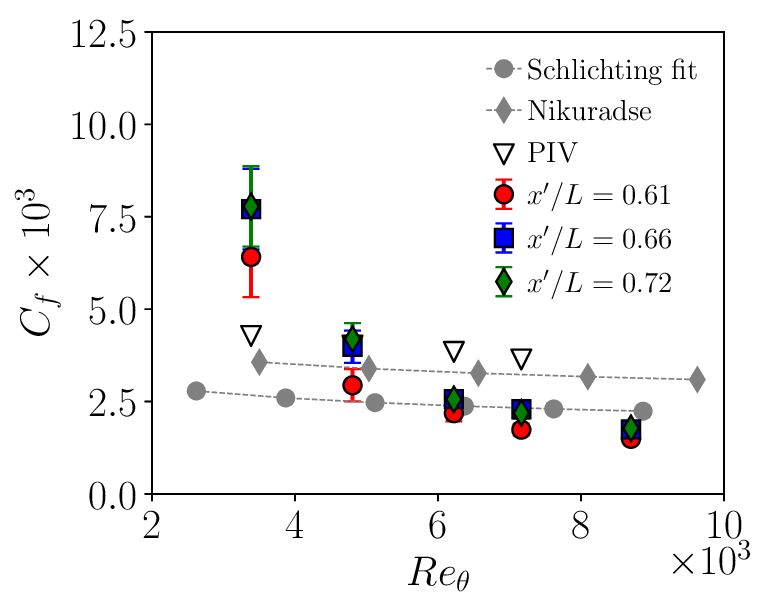}
        \caption{}
        \label{Fig_cf_comparison}
    \end{subfigure}
    \hspace{0.04\textwidth}  
    \begin{subfigure}[t]{0.31\textwidth}
        \centering
        \includegraphics[width=\textwidth]{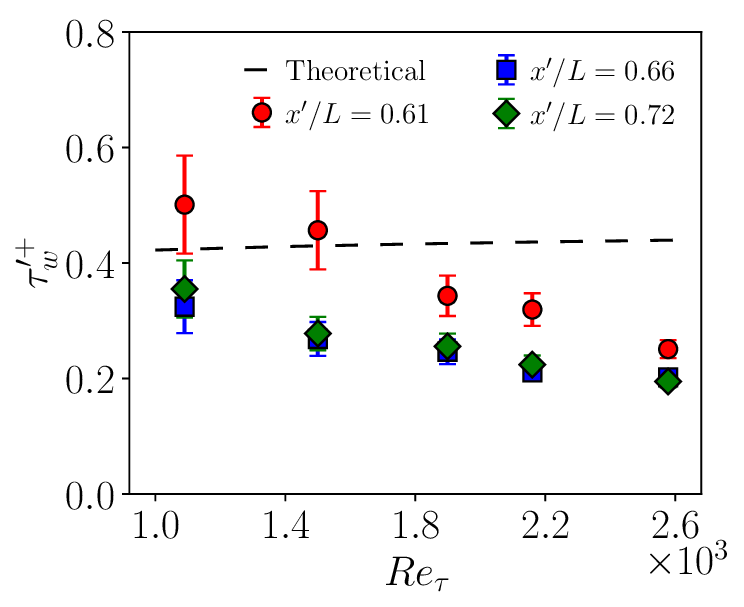}
        \caption{}
        \label{Fig_tau_prime_comparison}
    \end{subfigure}
    \caption{Comparison of turbulent quantities with literature under ZPG conditions for different sensors. (a) Friction coefficient ($C_f$) vs. momentum Reynolds number ($Re_{\theta}$) and (b) normalized wall shear stress fluctuations ($\tau_w'^{+}$) vs. friction Reynolds number ($Re_{\tau}$).}    
    \label{Fig_Cf_tau_rms_ZPG}
\end{figure}

This section examines $Re_\tau$ effects on direct wall shear stress measurements in a canonical ZPG TBL. The analysis focuses on the skin friction coefficient ($C_f = \tau_w / (0.5 \rho U_\infty^2)$) and the normalized fluctuation magnitude of wall shear stress ($\tau_w'^+=\tau'_{w,\mathrm{rms}}/\overline{\tau_w}$), which are shown in figure \ref{Fig_Cf_tau_rms_ZPG} as function of Reynolds number. Both $C_f$ and $\tau_w'^+$ (colored markers) exhibit spatial variations along the streamwise direction. The first sensor records a slightly lower $C_f$ and therefore a larger $\tau_w'^+$ across the explored Reynolds number range compared to the two downstream sensors, which generally show better agreement.  

Figure \ref{Fig_cf_comparison} shows the variation of $C_f$ with $Re_{\theta}$, comparing direct measurements (colored markers) to reference curves from \citet{Nagib2007ApproachLayers}, the Schlichting fit (dashed line with circles) and Nikuradse values (dashed line with diamonds). Also reported are estimates based on 2D PIV velocity fields acquired in the same facility \citep{DatabasePIV2023}, using the Clauser chart method with $\kappa=0.41$ and $\mathrm{B}=5.0$. The colored markers each indicate a sensor at a different streamwise location in the present experimental setup. Figure \ref{Fig_cf_comparison} is plotted against $Re_\theta$ to be consistent with the reference data, while figure \ref{Fig_tau_prime_comparison} is plotted against $Re_\tau$ to be consistent with the theory that it is compared against. The general trend that $C_f$ decreases with increasing $Re_\tau$ is consistent between the direct measurements and the reference data in figure \ref{Fig_cf_comparison}. However, the direct measurements show some disagreement with the reference data for both low and high $Re_\theta$. $C_f$ is overestimated at the lowest Reynolds number and underestimated at higher Reynolds numbers. While a very large discrepancy between direct measurements and literature data is observed at the lowest $Re_{\theta}$, this decreases to less than 20\% at the largest $Re_{\theta}$ when comparing to the Schlichting fit. At low flow speeds, the discrepancy may be attributed to a low signal-to-noise ratio in the experimental setup, which contributes to significant uncertainty, as indicated with the error bars. In the range $4804\leq Re_{\theta} \leq 7166$, the measured skin friction coefficients converge toward the literature values across all sensors, with better agreement as Reynolds number increases. This trend implies the existence of a specific interval of Reynolds numbers where the difference between measured and reference data reaches a minimum. This behavior may result from a balance between experimental signal-to-noise ratio and sensor spatial and temporal resolutions. Beyond this range, the measurements follow the expected trend but remain slightly smaller than literature values. The discrepancy between literature and experimental values may also be partially attributed to the absence of in-situ sensor calibration. Future experiments incorporating in-situ calibration could help further reduce discrepancies.

Figure \ref{Fig_tau_prime_comparison} shows the variation of $\tau'^+_{w}$ with $Re_{\tau}$ for the three sensors alongside the theoretical prediction $\tau'^+_{w} = 0.298 + 0.018 \ln Re_{\tau}$ proposed by \cite{OrluSchlatter2011}, which yields expected values between 0.42 to 0.44 within the explored $Re_{\tau}$ range. Generally, the raw wall shear stress and its fluctuations increase with $Re_\tau$, with $\tau_{w,\mathrm{rms}}$ increasing by 49.9\% between the lowest and highest $Re_{\tau}$. However, the normalized fluctuations $\tau_w'^+$ decrease with increasing $Re_{\tau}$. The upstream sensor ($x'/L=0.61$) records higher $\tau_w'^+$ values than the two downstream sensors across the entire $Re_{\tau}$ range, exceeding theoretical prediction for $Re_{\tau} \leq 1500$ due to a lower measured mean wall shear stress. All three sensors exhibit a decreasing trend in $\tau_w'^+$ with $Re_{\tau}$, in contrast to the weakly increasing logarithmic  prediction. This behavior has also been reported for thermal anemometry techniques and is attributed to resolution limitations. Prior studies in thermal anemometry \citep{Alfredsson1988, Osterlund1999ExperimentalFlow, Colella2003, OrluSchlatter2011} have shown that as $Re_{\tau}$ increases, the inner-scaled sensor dimension grows, leading to amplified spatial averaging and attenuation of measured fluctuations.

\subsection{FAPG influence on mean and fluctuating skin friction}
\begin{figure}[htbp]
    \centering
    \begin{subfigure}[t]{0.3\textwidth}
        \centering
        \includegraphics[width=\textwidth]{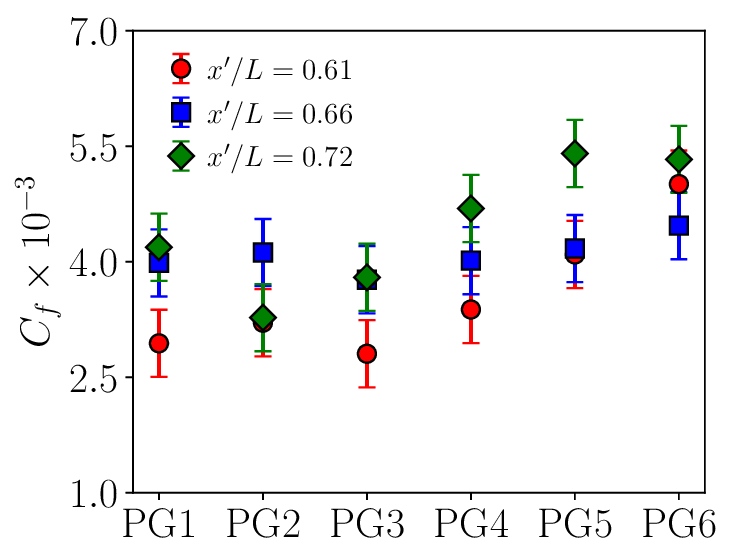}
        \caption{$C_f$ at $Re_{\tau} = 1500$}
        \label{Fig_Cf_Re_1500}
    \end{subfigure}
    \hspace{0.04\textwidth}  
    \begin{subfigure}[t]{0.3\textwidth}
        \centering
        \includegraphics[width=\textwidth]{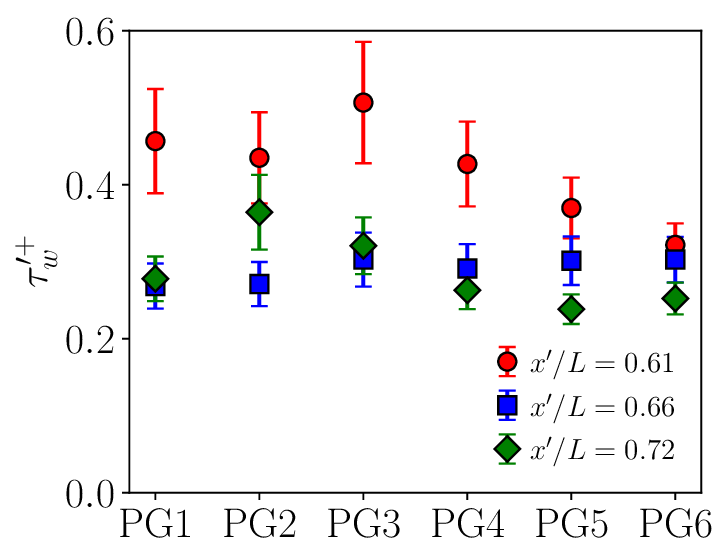}
        \caption{$\tau_w'^+$ at $Re_{\tau} = 1500$}
        \label{Fig_tau_rms_Re_1500}
    \end{subfigure}
    \vskip\baselineskip
    \begin{subfigure}[t]{0.3\textwidth}
        \centering
        \includegraphics[width=\textwidth]{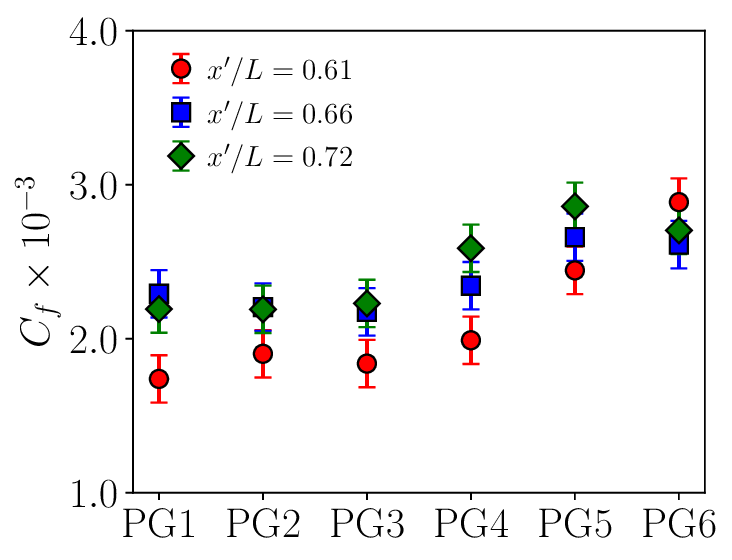}
        \caption{$C_f$ at $Re_{\tau} = 2160$}
        \label{Fig_cf_Re_2160}
    \end{subfigure}
    \hspace{0.04\textwidth}  
    \begin{subfigure}[t]{0.3\textwidth}
        \centering
        \includegraphics[width=\textwidth]{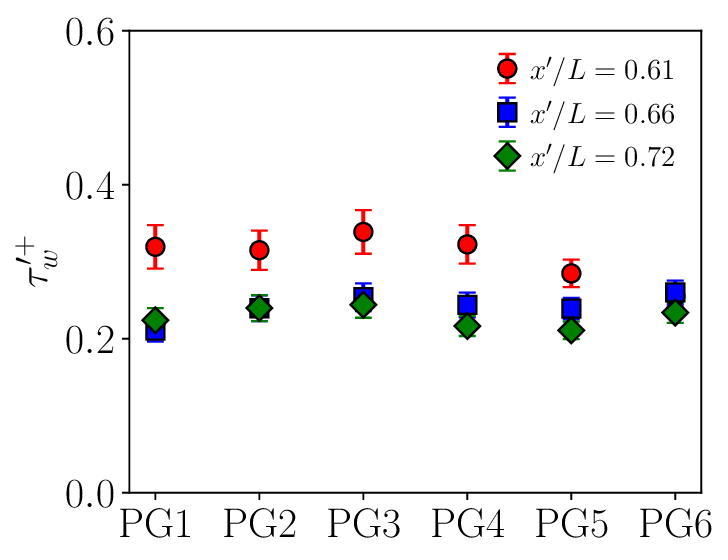}
        \caption{$\tau_w'^+$ at $Re_{\tau} = 2160$}
        \label{Fig_tau_rms_Re_2160}
    \end{subfigure}
    \caption{$C_f$ and $\tau_w'^+$ as functions of FAPG strength ($\mathrm{PG1-PG6}$) for two Reynolds numbers: (a)-(b) $Re_{\tau} = 1500$, (c)-(d) $Re_{\tau} = 2160$.}
    \label{Fig_cf_tau_rms_re_1500_2160}
\end{figure}

Figure \ref{Fig_cf_tau_rms_re_1500_2160} shows $C_f$ and $\tau_w'^{+}$ as functions of the FAPG strength ($\text{PG}i,\ i=1,...,6$) for two Reynolds numbers: $Re_{\tau}=1500$ in figures \ref{Fig_Cf_Re_1500} and \ref{Fig_tau_rms_Re_1500}, and $Re_{\tau}=2160$ in figures \ref{Fig_cf_Re_2160} and \ref{Fig_tau_rms_Re_2160}. At $Re_{\tau} = 1500$, both quantities exhibit pronounced variations across PG sequences. At the most upstream and downstream positions ($x'/L = 0.61$ and $0.72$), a stronger FAPG sequence leads to increased $C_f$. This is an expected trend, even though the local pressure gradient is adverse, because the upstream favorable pressure gradient yields an overall accelerated boundary layer in this region \citep{parthasarathy2023family}. An inverse correlation between $C_f$ and $\tau_w'^{+}$ is observed, with sensor-specific variations. Strong FAPG sequences show reduced $\tau_w'^{+}$ for the most upstream and downstream positions. Minimal variations are observed in the data of the middle sensor ($x'/L=0.66$). At the highest Reynolds number, the coefficient of friction is observed to increase with FAPG strength, as expected, for all three sensors. At this Reynolds number, there is a nearly constant $\tau_w'^{+}$, independent of FAPG strength. However, this trend may be limited by attenuation effects.

\begin{figure}[htbp]
    \centering
    %%%%%%%%%%%%%%%%%%%%%%%%
    % First row
    \begin{subfigure}[b]{0.31\textwidth}
        \includegraphics[width=\textwidth]{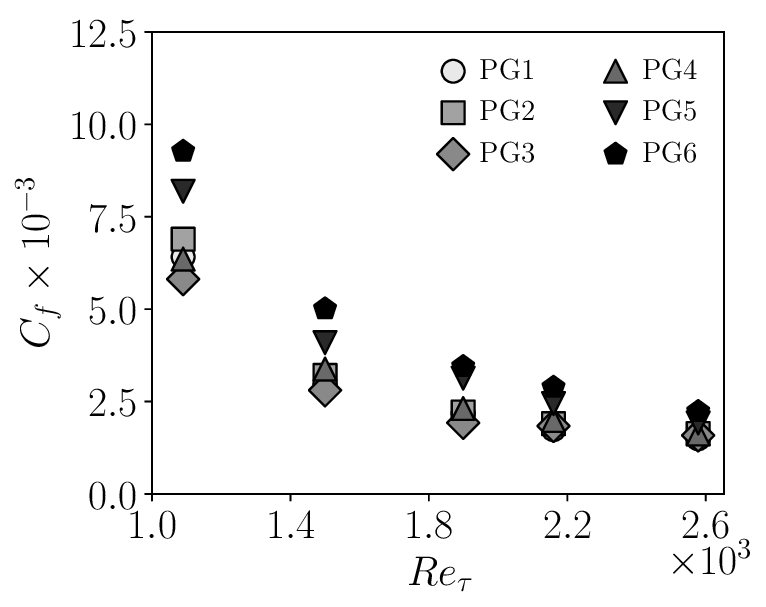}
        \caption{$C_f$ at $x'/L = 0.61$}
        \label{Fig_cf_x1}
    \end{subfigure}
    \hfill
    \begin{subfigure}[b]{0.31\textwidth}
        \includegraphics[width=\textwidth]{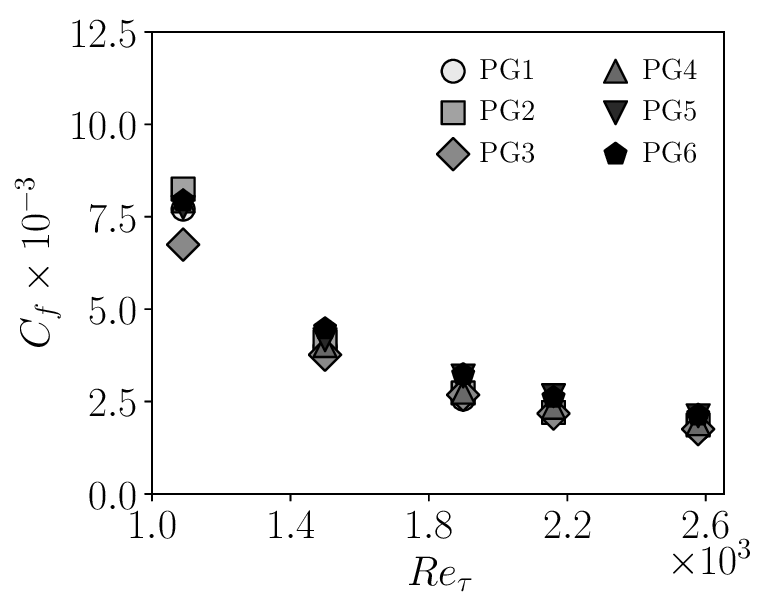}
        \caption{$C_f$ at $x'/L = 0.66$}
        \label{Fig_cf_x2}
    \end{subfigure}
    \hfill
    \begin{subfigure}[b]{0.31\textwidth}
        \includegraphics[width=\textwidth]{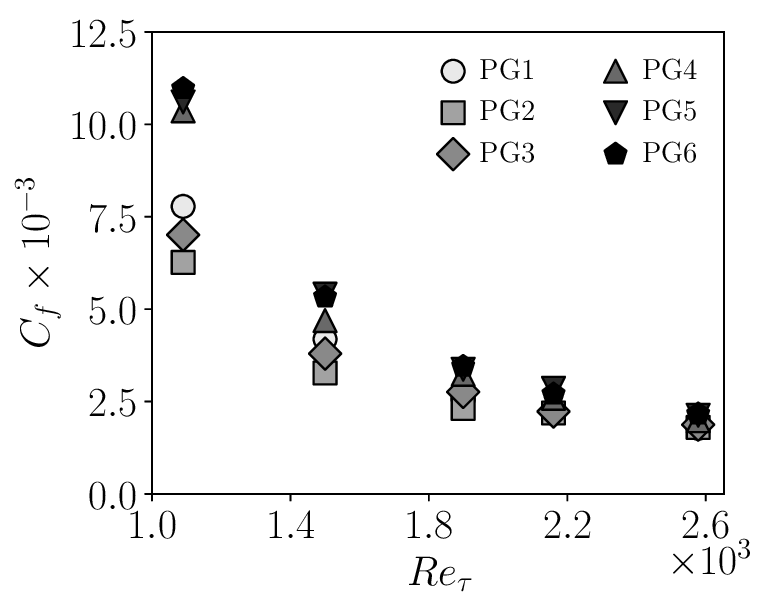}
        \caption{$C_f$ at $x'/L = 0.72$}
        \label{Fig_cf_x3}
    \end{subfigure}
    \vspace{1em}
    %%%%%%%%%%%%%%%%%%%%%%%%
    % Second row
    \begin{subfigure}[b]{0.3\textwidth}
        \includegraphics[width=\textwidth]{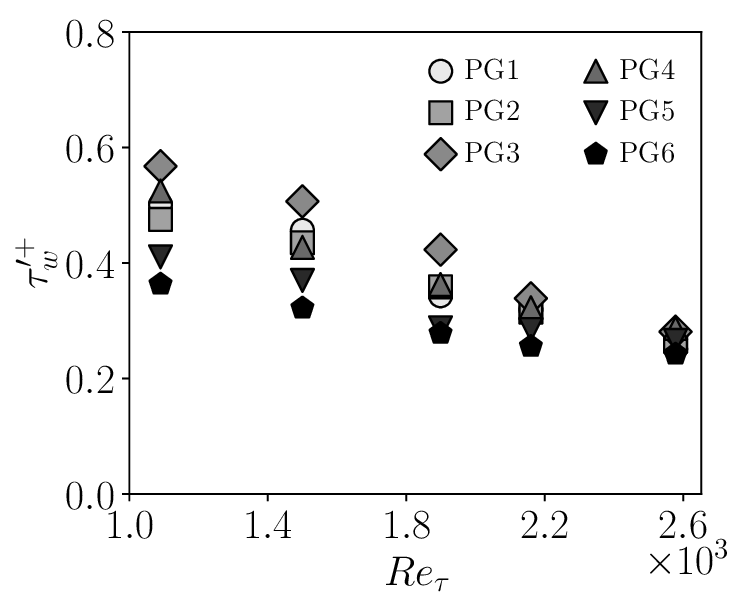}
        \caption{$\tau_w'^{+}$ at $x'/L = 0.61$}
        \label{Fig_tau_plus_x1}
    \end{subfigure}
    \hfill
    \begin{subfigure}[b]{0.3\textwidth}
        \includegraphics[width=\textwidth]{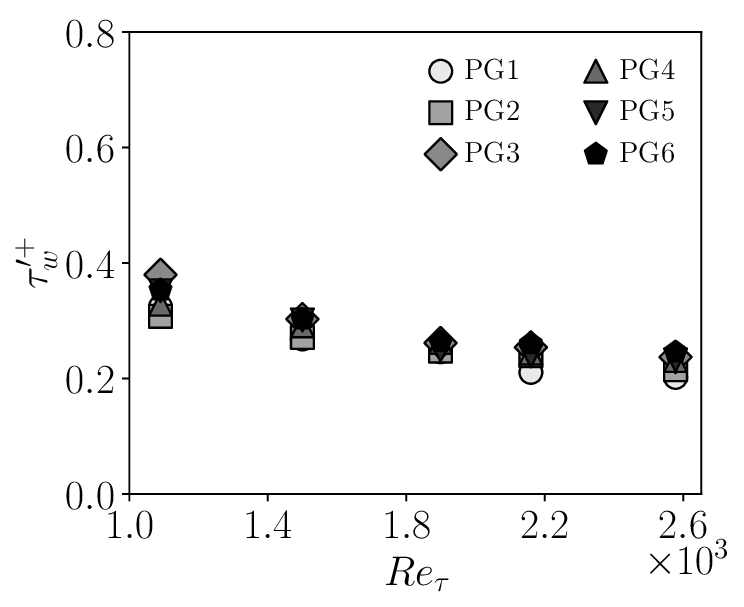}
        \caption{$\tau_w'^{+}$ at $x'/L = 0.66$}
        \label{Fig_tau_plus_x2}
    \end{subfigure}
    \hfill
    \begin{subfigure}[b]{0.3\textwidth}
        \includegraphics[width=\textwidth]{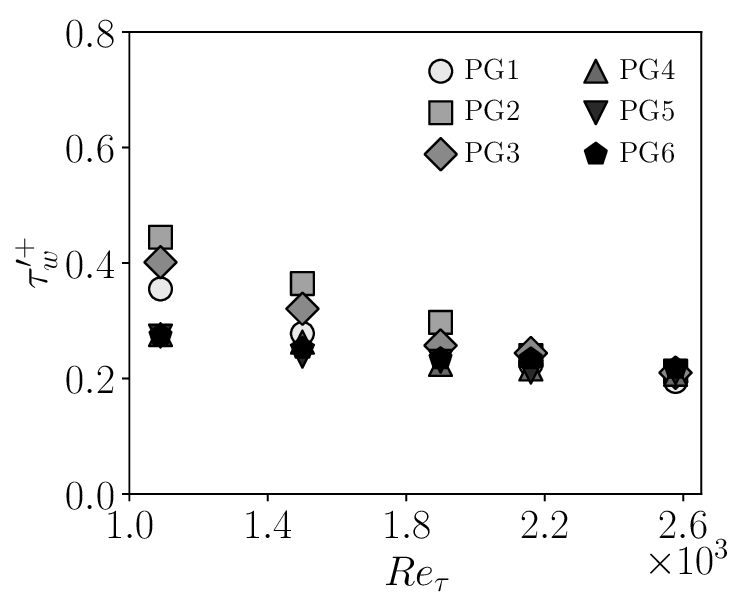}
        \caption{$\tau_w'^{+}$ at $x'/L = 0.72$}
        \label{Fig_tau_plus_x3}
    \end{subfigure}
    \caption{Variations of $C_f$ and $\tau_w'^+$ with $Re_{\tau}$ for different FAPG strengths (PG1–PG6) at different streamwise locations.}
    \label{Fig_re_cf_tau_rms_all}
\end{figure}

To further analyze the history effects induced by the FAPG sequences under different flow regimes, figure \ref{Fig_re_cf_tau_rms_all} presents $C_f$ and $\tau_w'^+$ as functions of $Re_{\tau}$ at the three sensor locations in the APG region: $x'/L = 0.61$ (figures \ref{Fig_cf_x1}, \ref{Fig_tau_plus_x1}), $x'/L = 0.66$ (figures \ref{Fig_cf_x2}, \ref{Fig_tau_plus_x2}), and $x'/L = 0.72$ (figures \ref{Fig_cf_x3}, \ref{Fig_tau_plus_x3}). The results highlight the influence of the FAPG strength, revealing its additional contribution to $C_f$ and $\tau_w'^{+}$ across different $Re_{\tau}$. The $C_f$ decreases with increasing $Re_{\tau}$ independently of the FAPG strength applied. For almost every Reynolds number and sensor location, the strongest FAPG yields the highest $C_f$. This may indicate the strong influence of the upstream FPG on the local skin friction in the APG region for the strongest two FAPG cases \citep{parthasarathy2023family}. The sensitivity of $C_f$ to the pressure gradient drops with increasing $Re_\tau$, with smaller variability across PGs. The weaker FAPG strengths show more complicated results. For the farthest upstream case, PG3 shows the minimum $C_f$ for the first three $Re_\tau$ cases. This may indicate that the local APG is able to dominate over the upstream FPG for these cases. This same trend is observed in the mid-location signature for the first two Reynolds numbers. In the farthest downstream case, both PG2 and PG3 show lower $C_f$ than PG1, which may indicate that the local APG is able to be more dominant than the FPG effects. For the farthest downstream case, perhaps its location farther downstream is related to its increased APG-like behavior for two rather than only one of the weaker FAPG cases. 

The trends observed for the mean skin friction in figures \ref{Fig_cf_x1}, \ref{Fig_cf_x2}, and \ref{Fig_cf_x3} are consistent with those observed for the root-mean-square of the skin friction in figures \ref{Fig_tau_plus_x1}, \ref{Fig_tau_plus_x2}, and \ref{Fig_tau_plus_x3}. Table \ref{tab:cf_uncertainty} reports the relative uncertainty for the skin friction coefficient associated to each sensor for each different flow condition. While attenuation effects impact the resolution of turbulent fluctuations for $Re_{\tau} \geq 1500$, general trends with respect to FAPG strength can still be inferred. Fluctuation amplitude reduces with increasing Reynolds number, regardless of pressure gradient, though this is a stronger dependence for the weaker pressure gradients. The fluctuation of the skin friction is a stronger function of pressure gradient for lower Reynolds numbers. The strongest two FAPG cases show the lowest fluctuation amplitudes for all Reynolds numbers, except for the mid-location sensor, but the sensitivity is so low that there is significant uncertainty (see table \ref{tab:cf_uncertainty}) that prevents full conclusions for the mid-location sensor. The observed reductions in local skin friction fluctuations are consistent with the upstream FPG effects dominating the local skin friction, even in the APG region. Again, the weaker pressure gradients show more complex results for the fluctuations in figures \ref{Fig_tau_plus_x1}, \ref{Fig_tau_plus_x2}, and \ref{Fig_tau_plus_x3}, just as they did for the mean conditions. PG3 shows the highest fluctuations for the farthest upstream case and PG2 and PG3 both show high fluctuations for the farthest downstream case. These higher fluctuations are likely indicative of the local APG being dominant over the upstream FPG for the local skin friction at the weaker FAPG conditions.  
 
\subsection{Large-scale turbulent structure influence on the wall shear stress}
\begin{figure}[htbp]
    \centering
    \begin{subfigure}[t]{0.32\textwidth}
        \centering
        \includegraphics[width=\textwidth]{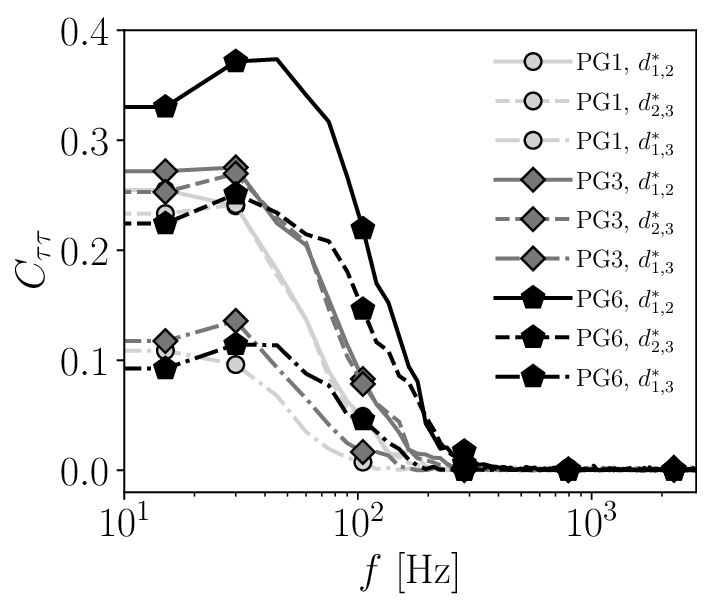}
        \caption{$Re_{\tau} = 1500$}
        \label{Fig_cohe_Re_1500}
    \end{subfigure}
    \hspace{0.04\textwidth}  
    \begin{subfigure}[t]{0.32\textwidth}
        \centering
        \includegraphics[width=\textwidth]{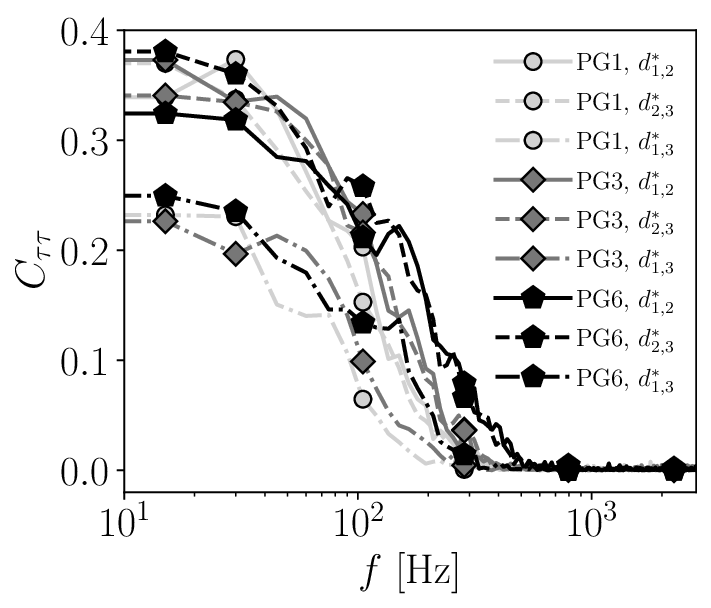}
        \caption{$Re_{\tau} = 2160$}
        \label{Fig_cohe_Re_2160}
    \end{subfigure}
    \caption{Coherence spectrum ($C_{\tau\tau}$) as a function of frequency $(f)$ under different FAPG sequences (PG1, PG3, PG6) for $Re_{\tau}=1500$ (a) with $d_{1,2}^* = d_{2,3}^* = 0.521$, $d_{1,3}^* = 1.042$ and $Re_{\tau}=2160$ (b) with $d_{1,2}^* = d_{2,3}^* = 0.578$, $d_{1,3}^* = 1.156$. Sensor pairs: solid for $d_{1,2}^*$, dashed for $d_{2,3}^*$, and dash-dotted for $d_{1,3}^*$.}
    \label{Fig_cohe_Re_1500_19}
\end{figure}

We next investigate the influence of large-scale turbulent structures on the wall shear stress as the structures pass through the pressure gradient sequence. It is known that large-scale turbulent structures leave an imprint on the wall-shear stress \citep{o2009chasing} over a broad range of frequencies, from tens of Hz up to the kilohertz \citep{Hutchins2011Three-dimensionalLayer, PabonThesis}. This large-scale structure frequency range is within the range that is fully resolved by the shear stress sensors. We expect the large-scale turbulent structures to be affected by the local history effects (e.g. likely still dominated by the upstream FPG), as large-scale motions in the outer layer of a TBL respond to pressure changes with a streamwise delay of approximately $10\delta$ \citep{gungor2024turbulent}, and the probes are positioned near the early stages of this delayed response region. To characterize the influence of large-scale structures on the shear stress, we analyze the coherence spectrum, mean cross-correlation, and convection velocity, which provide insights about the spatial and temporal scales of the flow. We will utilize filtering to focus on large-scale signatures.

The coherence spectrum of $\tau_w$ simultaneously measured at different streamwise locations is
\begin{equation}
C_{\tau\tau}(f)=\dfrac{|\Phi_{\tau_{ij}}(f)|^2}{\Phi_{\tau_{i}}(f)\Phi_{\tau_{j}}(f)},
\end{equation}
which quantifies the correlation in the frequency domain between two time-series measurements. Here, $\Phi_{\tau_{ij}}(f)$ is the cross spectral density of $\tau_w$ signals for sensors $i$ and $j$ for pairs $i=1,2$, $j=2,3$, with $i\neq j$, while $\Phi_{\tau_{i}}(f)$ and $\Phi_{\tau_{j}}(f)$ are the auto-spectral density estimates of the signals acquired at sensors $i$ and $j$. Figure \ref{Fig_cohe_Re_1500_19} reports the coherence spectrum for two different Reynolds numbers, $Re_{\tau}=1500$ (figure \ref{Fig_cohe_Re_1500}) and $Re_{\tau}=2160$ (figure \ref{Fig_cohe_Re_2160}). The results are shown as a function of the frequency to allow direct comparison with previous work \citep{PabonThesis}. For each case, $C_{\tau\tau}(f)$ is shown for different sensor pairs ($d_{1,2}^*,\ d_{2,3}^*,\ d_{1,3}^*$) under FAPG sequences of varying magnitudes (PG1, PG3, PG6).  

The coherence spectrum for $Re_\tau = 1500$ indicates the highest coherence at low frequencies with very low values of the coherence spectrum at higher frequencies. Note that $d^*=0.52$ here, suggesting that wall shear stress signatures with a temporal frequency of over approximately 200 Hz are not highly correlated past half a boundary layer thickness in spatial convection. As expected, coherence decreases with increasing sensor spacing, with the coherence spectrum with a spacing of $d_{1,3}^*$ exhibiting lower coherence than those captured by $d_{1,2}^*$ and $d_{2,3}^*$, regardless of the FAPG strength. For the ZPG case (PG1), coherence approaches zero at approximately 195 Hz ($f\delta/\uinf \approx 0.8$) for spacings of $d_{1,2}^* = d_{2,3}^*=0.521$, and around to 390 Hz ($f\delta/\uinf \approx 1.5$) for motions larger than $d_{1,3}^*=1.042$. Increasing the FAPG strength leads to larger coherence at higher frequencies across all sensor pairs. Larger coherence levels may suggest elongated turbulent structures or a larger influence of the structures on the wall. Also, higher temporal frequencies being associated with a fixed spatial wavenumber suggests higher convective speeds. For PG3, the increase is modest without differences of coherence across the upstream ($d_{1,2}^*$) and downstream ($d_{2,3}^*$) sensor pairs. Coherence decays to zero around 250 Hz ($f\delta/\uinf \approx 1.0$) for both $d_{1,2}^*$ and $d_{2,3}^*$. In contrast, PG6 exhibits more pronounced difference between upstream $d_{1,2}^*$ and downstream $d_{2,3}^*$ coherence, with the upstream pair shower significantly higher values at low frequencies than either the downstream pair or the ZPG case. The peak coherence reaches $C_{\tau\tau}\approx0.37$ between 30-50 Hz for $d_{1,2}^*$, representing a 48\% increase over the PG1 case. The significant increase in the spectral coherence at low frequencies may imply a higher probability of longer structures affecting the wall shear stress, which would be consistent with the influence of a FPG. By contrast, at higher frequencies both the upstream and downstream pairs show higher coherence, which may reflect an increase in convective velocity of the turbulent structure's signature on the wall shear stress. 
 
At $Re_{\tau}=2160$, coherence spectrum reflect an overall increase in high-frequency content across all sensor pairs, consistent with the shift towards higher frequency energy content at higher Reynolds numbers. The zero-coherence frequency increases from 195 Hz at $Re_{\tau}=1090$ to around 405 Hz ($f\delta/U_{\infty} \approx 0.9$) for $d_{1,2}^*$, and similar values are reached for $d_{2,3}^*$.  The influence of sensor spacing remains consistent, with shorter pairs yielding higher coherence. However, at higher Reynolds number, the effects of the FAPG sequence are less pronounced, suggesting a reduced sensitivity to PG effects. Along with this reduction in sensitivity to PG, the significant spatial variation between the upstream and downstream pairs for the strongest pressure gradient case is suppressed and reversed at the higher Reynolds number. This is somewhat consistent with the trends observed in $\tau'^+_w$; at lower Reynolds number, the upstream sensor shows significantly higher variability than the downstream sensor in figure \ref{Fig_tau_rms_Re_1500}, while a much smaller difference is observed at the higher Reynolds number in figure \ref{Fig_tau_rms_Re_2160}.

We also consider the two-point cross-correlation coefficient of $\tau_w$, which provides information about the relative similarity of the two signals and the time delay associated with an effective convection speed of the wall shear stress fluctuations.
\begin{equation}
    \mathrm{R}(\tau_i,\tau_j,\Delta\tau) = \dfrac{\mathrm{mean}(\tau(s_i,t)\cdot \tau(s_j,t+\Delta \tau))}{\sigma(\tau(s_i,t))\ \sigma(\tau(s_j,t))},
\end{equation}
where $\tau(s_i,t)$ and $\tau(s_j,t)$ are the $\tau_w$ time-series at sensors $i$ and $j$ at time $t$, $\sigma(\tau'(s_i,t))$ and $\sigma(\tau'(s_j,t))$ are their respective standard deviations. The time lag $\Delta\tau$ corresponds to the shift that maximizes the correlation. Indices are defined as $i=1,2$, and $j=i+1$. The streamwise correlation of $\tau_w$ is used to estimate a convection velocity, given by $ U_c = \kappa \frac{\Delta x_s}{\Delta\tau}$, where $\Delta x_s = 0.025$ is the sensor spacing and $\kappa = 1, 2$ is the probe separation factor. Cross-correlation and convection velocity estimates may be affected by high-frequency noise in the signals. To mitigate this disturbance, the shear stress signals are pre-processed using a third-order Butterworth low-pass filter. The process for selecting the adaptive cutoff frequency is provided in Appendix \ref{app:A}. 

\begin{figure}[htbp]
    \centering
    \begin{subfigure}[t]{0.3\textwidth}
        \includegraphics[width=\textwidth]{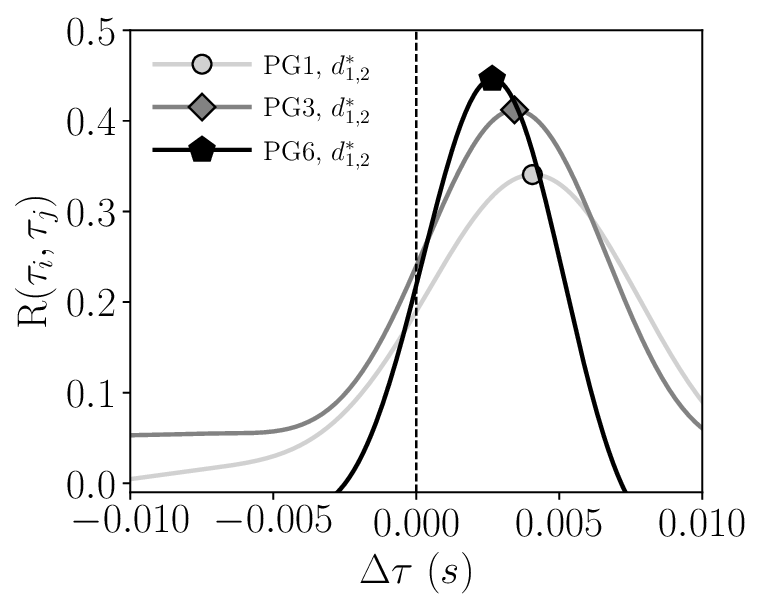}
        \caption{$Re_{\tau} = 1500$}
        \label{Fig_CC_Re_tau_1500}
    \end{subfigure}
    \hfill
    \begin{subfigure}[t]{0.3\textwidth}
        \includegraphics[width=\textwidth]{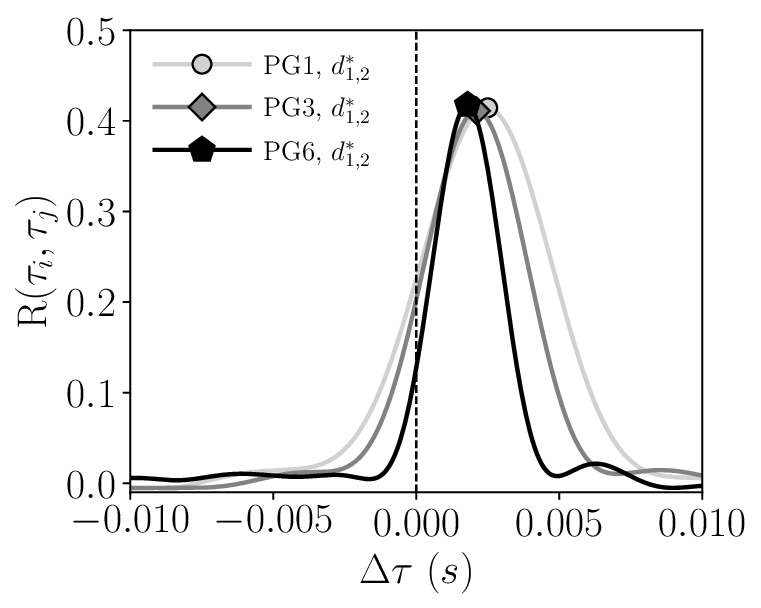}
        \caption{$Re_{\tau} = 2160$}
        \label{Fig_CC_Re_tau_2160}
    \end{subfigure}
    \hfill
    \begin{subfigure}[t]{0.298\textwidth}
        \includegraphics[width=\textwidth]{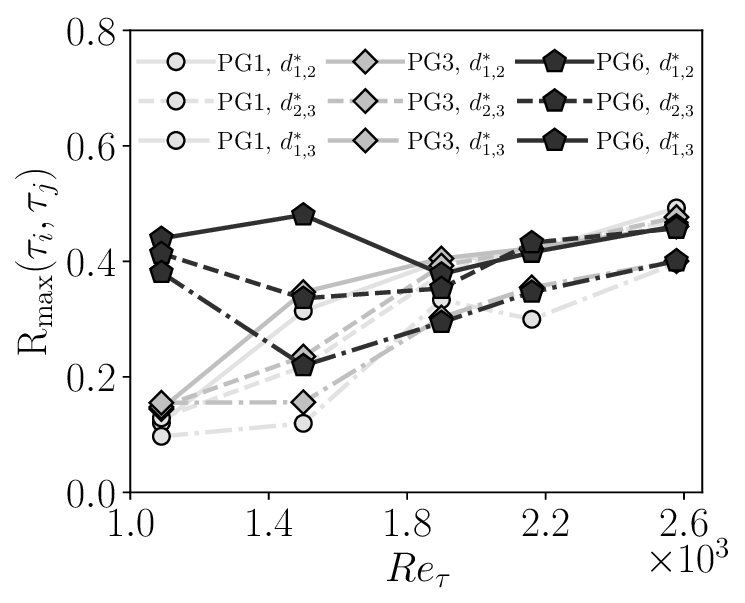}
        \caption{}
        \label{Fig_CC_All_Re_tau}
    \end{subfigure}
    \caption{Cross-correlation coefficient ($\mathrm{R}(\tau_i,\tau_j)$) as a function of the time shift ($\Delta \tau$) for the sensor pair $d_{1,2}^*$ under three different pressure gradients (PG1, PG3, PG6) at $Re_{\tau}=1500$ (a) and $Re_{\tau}=2160$ (b). And max value as a function of $Re_\tau$ and PG sequence (c).}
    \label{Fig_corr_coeffs_Re_1500_2160}
\end{figure}

After filtering the wall shear stress signals using the proposed adaptive strategy, $R(\tau_i,\tau_j)$ is evaluated in figure \ref{Fig_corr_coeffs_Re_1500_2160} as a function of time lag ($\Delta \tau$), following the approach in \citep{Hutchins2011Three-dimensionalLayer} to enable direct comparison. Results are reported for the sensor pair $d_{1,2}^*$ under three FAPG (PG1, PG2, PG3) at two Reynolds numbers, for one of the three independent acquisitions. The positive time lags ($\Delta\tau>0$) confirm the downstream convection of wall-shear stress perturbations. Both the freestream condition and FAPG strength affect $\Delta\tau$.  At $Re_{\tau}=1500$ (figure \ref{Fig_CC_Re_tau_1500}), slightly broader correlation peaks are visible, implying a wider range of time shifts over which structures retain correlation. The opposite occurs at $Re_{\tau}=2160$ (figure \ref{Fig_CC_Re_tau_2160}), where the correlation becomes more localized in time and is associated with sharper peaks and shorter $\Delta \tau$. The effects of FAPG strength are more pronounced at low Reynolds numbers and, while the results in figure \ref{Fig_corr_coeffs_Re_1500_2160} correspond to the sensor pair $d_{1,2}^*$, similar trends but with non-monotonic responses are observed for other sensor pairs, in accordance with trends seen in the friction coefficient. Overall, increasing FAPG strength leads to higher peak correlation values and reduced $\Delta \tau$ across Reynolds numbers. Since an FPG accelerates the flow and stretches the elongation of large-scale structures in the spanwise and streamwise directions \citep{VOLINO2020108717}, the observed reduction in $\Delta \tau$ confirms the influence of the upstream FPG into the APG. As a result, the persistence of these elongated structures combined with the delayed response of the flow to pressure changes, leads to increased correlation across the sensor array.  

To generalize these findings across different flow conditions, figure \ref{Fig_CC_All_Re_tau} presents the average peak values of $R(\tau_i,\tau_j)$ across the three independent acquisitions as a function of $Re_{\tau}$ for all sensor pairs and FAPG sequences. The coefficient generally increases linearly with Reynolds number, except under the strongest FAPG sequence (PG6), which exhibits a non-monotonic trend. Across all conditions, $R(\tau_i,\tau_j)$ increases with FAPG strength and decreases with sensor spacing, with the largest separation ($d^*_{1,3}$) consistently showing the lowest values. 

\begin{figure}[htbp]
    \centering
    \begin{subfigure}{0.32\textwidth}
        \centering
        \includegraphics[width=\textwidth]{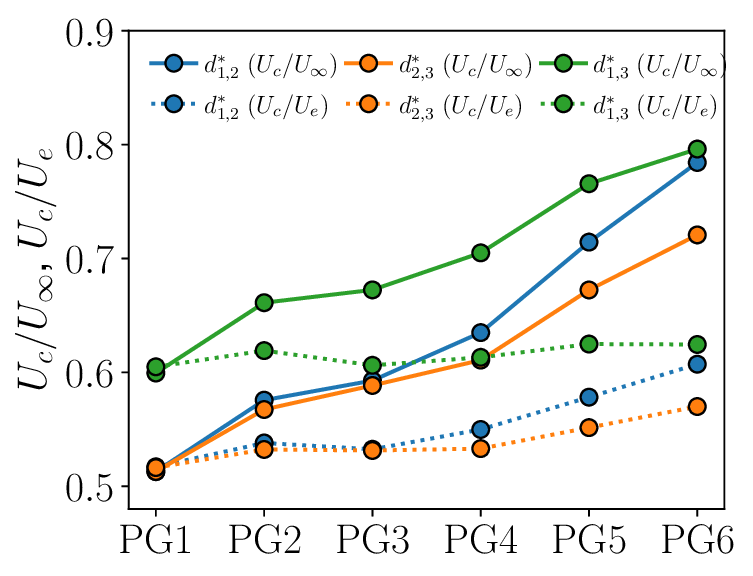}
        \caption{}
        \label{Fig_Uc_Ue_Re_1500}
    \end{subfigure}
    \hspace{0.04\textwidth}  
    \begin{subfigure}{0.32\textwidth}
        \centering
        \includegraphics[width=\textwidth]{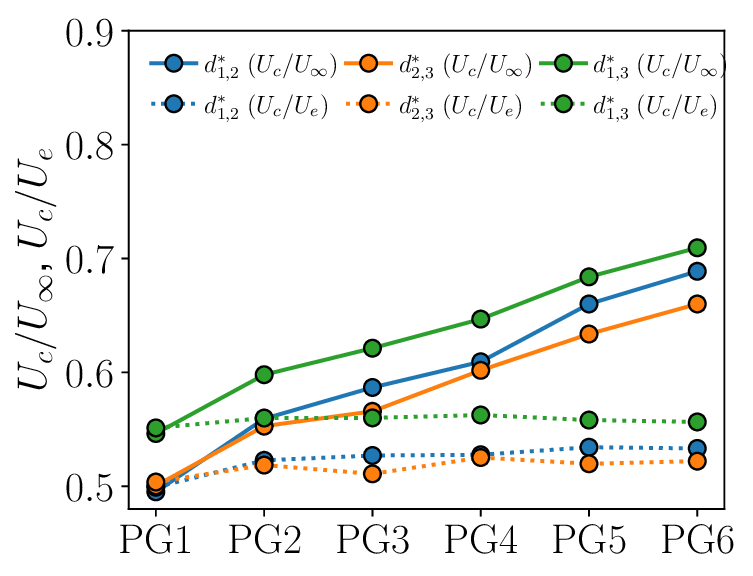}
        \caption{}
        \label{Fig_Uc_Ue_Re_2160}
    \end{subfigure}
    \caption{Normalized convection velocity for different sensor pairs ($d_{1,2}^*,\ d_{2,3}^*,\ d_{1,3}^*$) as a function of FAPG strength ($\mathrm{PG}_i,i=1,...,6$) at  $Re_{\tau}=1090$ (a) and $Re_{\tau}=2578$ (b)}
    \label{Fig_Uc_Ue_Re_1500_2160}
\end{figure}

The convection velocity results are displayed in figure \ref{Fig_Uc_Ue_Re_1500_2160}, where $U_c$ is normalized by the freestream velocity ($U_\infty$) and the edge velocity ($U_e$) for two Reynolds numbers, $Re_{\tau} = 1500$ (figure \ref{Fig_Uc_Ue_Re_1500}) and $Re_{\tau} = 2160$ (figure \ref{Fig_Uc_Ue_Re_2160}).  The edge velocity is extracted from 2D PIV measurements \citep{parthasarathy2023family} corresponding to the lowest Reynolds number and the same PG sequences reported in table \ref{tab:pg_conds} and extrapolated to free-stream speeds. For each sensor pair, the corresponding edge velocity is defined as the average $U_e$ at the two sensor locations.  For the ZPG case, the two normalizations yield similar values and the resulting normalized convection velocity values fall close to the range 0.56 to 0.83 \citep{willmarth1962measurements} for similar sensor spacings. For the non-zero pressure gradient cases, the choice of normalization significantly affects the interpretation of convection velocity trends across different FAPG. When normalized by $U_{\infty}$, $U_c/U_{\infty}$ linearly increases with FAPG strength for both $Re_\tau = 1500$ and $2160$, while when normalized by the local $U_e$, convection velocities vary much less with pressure gradient.

At $Re_\tau = 1500$ (figure \ref{Fig_Uc_Ue_Re_1500}), the sensor pairs $d_{1,2}^*$ and $d_{2,3}^*$ have normalized spacing of $0.521 \delta$, while the pair $d_{1,3}^*$ has a normalized spacing of $1.042 \delta$. For all pairs, $U_c/U_e$ remains nearly constant up to the PG4 sequence, after which it increases linearly with FAPG strength. The normalization with $U_e$ yields consistent results between $d_{1,2}^*$ and $d_{2,3}^*$, despite their streamwise separation, with $d_{1,2}^*$ showing larger values for PG4 to PG6. Additionally, larger-scale motions ($1.042 \delta$) captured by $d_{1,3}^*$ exhibit larger $U_c/U_e$ values, in agreement with previous studies \citep{Drozdz2023ConvectionGradient}.  \citet{willmarth1962measurements} also reported larger convection velocities when only low-frequency components were correlated. This trend aligns with the coherence results at this Reynolds number (figure~\ref{Fig_cohe_Re_1500}), where low-frequency content dominates. At $Re_\tau = 2160$ (figure \ref{Fig_Uc_Ue_Re_2160}), $U_c/U_e$ shows minimal variation across FAPG sequences of increasing strength. Although $d_{1,3}^*$ records slightly larger values of $U_c/U_e$ than the single spaced probes, the differences are smaller than at $Re_\tau = 1500$, indicating a weaker sensitivity to spatial separation. While previous studies have reported increasing convection velocity with freestream velocity, this behavior is not observed in our data. However, this trend is consistent with the coherence results (figure \ref{Fig_cohe_Re_2160}), which reveal diminished flow history effects from the FAPG sequence at high Reynolds number.

\section{Conclusions}\label{sec_conc}
In the first part of this study, we evaluated the performance of commercially available capacitive wall shear stress sensors across different Reynolds numbers ($Re_\tau = 1500$–$2160$) in a canonical ZPG TBL over a flat plate. We identified reasonable agreement between the measured mean and standard deviation with prior data and theory, though with spatial and temporal attenuation effects from the sensors' resolution. 
While $C_f$ decreased with increasing $Re_{\tau}$ as expected, the measurements slightly differed from reference data, with $C_f$ over-predicted at low $Re_{\tau}$ and under-predicted at higher $Re_{\tau}$. Similarly, $\tau_w'^+$ exhibited a decreasing trend opposite to theoretical predictions, due to attenuation.

The effects of five spatially varying FAPG of increasing strength on wall shear stress were assessed by directly measuring $\tau_w$ in the APG region downstream of the FPG. Results revealed strong flow history, with non-monotonic increases in $C_f$ and reductions in $\tau_w'^+$ as FAPG strength increased, with the latter possibly influenced by attenuation effects. Coherence spectrum highlighted a dominance of low-frequency, large-scale motions and increasing coherence with stronger FAPG. A dynamic low-pass filtering method based on the coherence results was introduced to isolate coherent features. Cross-correlation confirmed the downstream convection of wall-shear stress perturbations. Normalization of convection velocity with the edge velocity revealed an approximately constant $U_c/U_e$ across the sensor array. Limited sensor spacing effects were observed for $U_c/U_e$  at $Re_\tau=2160$. These findings confirmed the persistence of flow history effects on wall shear stress induced by pressure gradient sequences in non-equilibrium turbulent boundary layer and highlighted their Reynolds number dependence. 

\begin{appendices}
\section{}
\label{app:A}
The selection of the adaptive cutoff frequency ($f_{\text{cutoff}}$) is based on the coherence spectrum and involve two steps. First, for each flow condition and sensor pair, a dynamic upper frequency limit ($f_{\max,\theta_1}$) is defined as the highest frequency at which the coherence remains above a fraction ($\theta_1$) of its maximum, $ f_{\max,\theta_1} = \max \left\{f:C_{\tau\tau}(f)\geq \theta_1 \cdot \max(C_{\tau\tau})\right\}$. Second, the normalized cumulative coherence integral is used to determine $f_{\text{cutoff}}$:
\begin{equation}
    I(f) = \int_{f_{\min}}^{f_{\text{cutoff}}} C_{\tau\tau}(f') \, df', \qquad  \theta_2 = \frac{I(f_{\text{cutoff}})}{I(f_{\max,\theta_1})}
    \label{eq:c_integral}
\end{equation}
where $f_{\min}$ represents the lower bound of the frequency range. Here, $f_{\text{cutoff}}$ is identified such that $\theta_2$ reaches a prescribed fraction of the total coherence energy, retaining the dominant large-scale content while minimizing contributions from weakly correlated high-frequency content. This two steps method automatically adjusts $f_{\text{cutoff}}$ to different spectral decay behaviors observed under varying flow conditions (freestream velocity and FAPG strength).  

A sensitivity analysis across $\theta_1 = (0.1, 0.2, 0.3, 0.4)$ and $\theta_2=(0.60, 0.80, 0.90, 0.95)$ revealed that $\theta_1 = 0.1$ and $\theta_2=0.95$ provide robust cutoff frequencies. The impact of these parameters is quantified using the metric $\Delta_{\theta} = ({R_{\theta_{i+1}} - R_{\theta_i}})/R_{\theta_i}\ [\%]$, where $R$ denotes the cross-correlation coefficient and $\theta_{i}$, $\theta_{i+1}$ are successive values of the threshold parameters. At higher $Re_\tau$ ($\approx2600$), increasing $\theta_1$ (e.g., from 0.1 to 0.4) results in variations in $R$ exceeding 10\%, regardless of FAPG strength. In contrast, at lower $Re_\tau$, the variation remains below 5\%. The sensitivity decreases as $\theta_1 \rightarrow 0.1$, suggesting better robustness for lower thresholds. Nevertheless, the relative variation in $R$ decreases with increasing $\theta_2$. The reduction in sensitivity becomes noticeable beyond $\theta_2=0.90$, with changes limited to less than 2\% when increasing $\theta_2$ from 0.90 to 0.95, independently of the freestream conditions. 

The proposed filtering strategy is independent of flow specific assumption and does not rely on Taylor's hypothesis, which requires prescribing a convection velocity to relate temporal and spatial scales. While a cutoff frequency can be estimated by assuming $U_c\approx 0.8 U_{\infty}$ \citep{smits1989comparison}, the validity of Taylor's hypothesis deteriorates in non-equilibrium turbulent boundary layers. In contrast, our approach determines $f_{\text{cutoff}}$ directly from the coherence spectrum, adapting to the actual decay behavior. Table \ref{tab:cutoff} compares cutoff frequencies for the ZPG condition and a sensor spacing of $\Delta x_s = 0.025$\,m. The estimations based on Taylor use $U_c = 0.8U_{\infty}$, while $\theta_1 = 0.1$ and $\theta_2 = 0.95$ are the input for the coherence-based approach. The results show that the proposed method yields cutoff frequencies 1.7 to 1.8 times higher than those estimated with the frozen turbulence hypothesis. 

\begin{table}[htbp]
\centering
\caption{Cutoff frequencies for the ZPG flow.}
\label{tab:cutoff}
\begin{tabular}{lccccc}
\toprule
Method & Case 1 & Case 2 & Case 3 & Case 4 & Case 5 \\
\midrule
$f_{\text{cutoff}}$ (Hz) & 70  & 112 & 154 & 182 & 224 \\
$f_{\text{Taylor}}$ (Hz) & 39  &  61 &  86 & 103 & 132 \\
\bottomrule
\end{tabular}
\end{table}
\end{appendices}

\newpage

\backmatter

\bmhead{Acknowledgements}
This work was supported by the Office of Naval Research through grant no. N00014-21-1-2648 and by the National Science Foundation through grant no. 2339665.
The authors would like to thank David Mills, Ph.D., Jared Anderson, and Matt Barus from IC2 - Precision Measurement Solutions for their support. 

\bmhead{Author Contributions} Both authors contributed to the design of the experiments. MM conducted the experimental campaign, collected and analyzed the data. The first draft was written by MM and reviewed from TS-F. Both authors read and approved the manuscript. Project supervision and funding acquisition were performed by TS-F.

\section*{Declarations}
\bmhead{Conflict of interest} The authors have no relevant financial or non-financial interest to disclose. The authors have no competing interest to declare that are relevant to the content of this article. The authors have no financial or proprietary interests in any material discussed in this article.

\bibliography{sn-bibliography}

\end{document}